\documentclass[universe,conferencereport,accept,moreauthors,pdftex,10pt,a4paper,nohypertex]{mdpi}

\nolinenumbers

\firstpage{1} 
\articlenumber{41}
\doinum{10.3390/universe3020041}
\pubvolume{3}
\pubyear{2017}
\copyrightyear{2017}
\history{}

\setitemize{parsep=6pt,itemsep=0pt,leftmargin=*,labelsep=5.5mm}
\setenumerate{parsep=6pt,itemsep=0pt,leftmargin=*,labelsep=5.5mm}
\setlist[description]{itemsep=0mm}

\newcommand{\newc}{\newcommand*}

\long\def\begincomment#1\endcomment{%
        \begingroup\sf\baselineskip12pt#1\endgroup}

\newc{\etal}{\textrm{et al.}} 
\newc{\eg}{\textrm{e.g.}} 
\newc{\ie}{\textrm{i.e.}}
\newc{\etc}{\textrm{etc}}
\newc\vs{\textrm{vs.}}
\newc{\cl}{\rm {C.L.}}

\newc{\ev}{\ensuremath{\,\mathrm{eV}}}
\newc{\kev}{\ensuremath{\,\mathrm{keV}}}
\newc{\mev}{\ensuremath{\,\mathrm{MeV}}}
\newc{\gev}{\ensuremath{\,\mathrm{GeV}}}
\newc{\tev}{\ensuremath{\,\mathrm{TeV}}}
\newc{\MeV}{\mev} 
\newc{\TeV}{\tev}
\newc{\invpb}{\ensuremath{/\text{pb}}}
\newc{\invfb}{\ensuremath{\,\text{fb}^{-1}}}
\newc\nb{\ensuremath{\,\mathrm{nb}}} \newc\pb{\ensuremath{\,\mathrm{pb}}} \newc\fb{\ensuremath{\,\mathrm{fb}}}
\newc\pc{\ensuremath{\,\mathrm{pc}}}
\newc\kpc{\ensuremath{\,\mathrm{kpc}}}
\newc\mpc{\ensuremath{\,\mathrm{Mpc}}}
\newc\ps{\ensuremath{\,\mathrm{ps}}} 
\newc\cmeter{\ensuremath{\,\mathrm{cm}}} 
\newc\meter{\ensuremath{\,\mathrm{m}}} 
\newc\kmeter{\ensuremath{\,\mathrm{km}}}
\newc\second{\ensuremath{\,\mathrm{s}}}
\newc\msecond{\ensuremath{\,\mathrm{ms}}}
\newc\nsecond{\ensuremath{\,\mathrm{ns}}}
\newc\psecond{\ensuremath{\,\mathrm{ps}}}

\newc{\chisqmin}{\ensuremath{\chi^2_{\mathrm{min}}}}
\newc{\Delchisq}{\ensuremath{\Delta\chi^2}}
\newc{\chisq}{\ensuremath{\chi^2}}
\newc{\like}{\ensuremath{\mathcal{L}}}

\newc\lsim{\ensuremath{\mathrel{\rlap{\lower4pt\hbox{\hskip1pt$\sim$}}\raise1pt\hbox{$<$}}}}
\newc\gsim{\ensuremath{\mathrel{\rlap{\lower4pt\hbox{\hskip1pt$\sim$}}\raise1pt\hbox{$>$}}}}
\newc{\VEV}[1]{\ensuremath{\langle #1 \rangle}}
\newc{\dl}{\ensuremath{\stackrel{\leftarrow}{D}}}
\newc{\dr}{\ensuremath{\stackrel{\rightarrow}{D}}}

\newc{\bcenter}{\begin{center}}    \newc{\ecenter}{\end{center}}
\newc{\bfl}{\begin{flushleft}}    \newc{\efl}{\end{flushleft}}
\newc{\bfr}{\begin{flushright}}    \newc{\efr}{\end{flushright}}

\newc{\bi}{\begin{itemize}}
\newc{\ei}{\end{itemize}}
\newc{\bed}{\begin{description}}
\newc{\eed}{\end{description}}
\newc{\ben}{\begin{enumerate}}
\newc{\een}{\end{enumerate}}

\newc{\be}{\begin{equation}}
\newc{\ee}{\end{equation}}
\newc{\bea}{\begin{eqnarray}}
\newc{\eea}{\end{eqnarray}}
\newc{\ra}{\rightarrow}

\newc{\alphas}{\ensuremath{\alpha_s}}
\newc{\alphatwo}{\ensuremath{\alpha_2}}
\newc{\alphaone}{\ensuremath{\alpha_1}}
\newc{\alphai}[1]{\ensuremath{\alpha_{#1}}}
\newc{\alphaem}{\ensuremath{\alpha_{\mathrm{em}}}}
\newc{\alphaeff}{\ensuremath{\alpha_{\mathrm{eff}}}}
\newc{\sineff}{\ensuremath{\sin \theta_{\mathrm{eff}}}}
\newc{\sinsqeff}{\ensuremath{\sin^2 \theta_{\mathrm{eff}}}}
\newc{\dalphahad}{\ensuremath{\Delta \alpha_{\mathrm{had}}}}
\newc{\yt}{\ensuremath{h_t}} \newc{\yb}{\ensuremath{h_b}} \newc{\ytau}{\ensuremath{h_{\tau}}}
\newc\mz{\ensuremath{M_Z}} 
\newc\mw{\ensuremath{m_W}}
\newc\mZ{\mz}        \newc\mW{\mw}
\newc\mhsm{\ensuremath{ m_{H_{\mathrm{SM}}}}}
\newc{\mtop}{\ensuremath{ m_t}}               \newc{\mtpole}{\ensuremath{ M_t}}
\newc{\mbottom}{\ensuremath{ m_b}} 
\newc{\mtau}{\ensuremath{ m_{\tau}}}
\newc{\mt}{\mtpole}
\newc{\mb}{\mbottom} 
\newc{\rgg}{\ensuremath{R_{h}(\gamma\gamma)}}
\newc{\rzz}{\ensuremath{R_{h}(ZZ)}}
\newc{\rtwogg}{\ensuremath{R_{h_2}(\gamma\gamma)}}
\newc{\rtwozz}{\ensuremath{R_{h_2}(ZZ)}}
\newc{\ronegg}{\ensuremath{R_{h_1}(\gamma\gamma)}}
\newc{\ronezz}{\ensuremath{R_{h_1}(ZZ)}}
\newc{\rsiggg}{\ensuremath{R_{h_\textrm{sig}}(\gamma\gamma)}}
\newc{\rsigzz}{\ensuremath{R_{h_\textrm{sig}}(ZZ)}}
\newc{\llbar}{\ensuremath{\ell\bar{\ell}}}
\newc{\tauptaum}{\ensuremath{ \tau^+\tau^-}}
\newc{\qqbar}{\ensuremath{ q\bar{q}}} \newc{\ppbar}{\ensuremath{ p\bar{p}}}
\newc{\bbbar}{\ensuremath{ b\bar{b}}} \newc{\ttbar}{\ensuremath{ t\bar{t}}}
\newc{\ffbar}{\ensuremath{ f\bar{f}}} \newc{\tautaubar}{\ensuremath{ \tau\bar{\tau}}}
\newc{\mchi}{\ensuremath{m_{\chi}}}
\newc{\squark}{\ensuremath{\tilde{q}}}
\newc{\slepton}{\ensuremath{\tilde{l}}}
\newc{\gluino}{\ensuremath{\tilde{g}}} 
\newc{\mgluino}{\ensuremath{{m_{\gluino}}}}
\newc{\tone}{\ensuremath{{\tilde{t}_1}}}

\newc{\sthw}{\ensuremath{ \sin\theta_W}}              \newc{\cthw}{\ensuremath{\cos\theta_W}}
\newc{\tanthw}{\ensuremath{ \tan\theta_W}}              \newc{\cotthw}{\ensuremath{\cot\theta_W}}
\newc{\ssqthw}{\ensuremath{\sin^2 \theta_W}}
\newc{\msbar}{\ensuremath{\overline{MS}}} \newc{\drbar}{\ensuremath{\overline{DR}}}
\newc{\mtmtsmmsbar}{\ensuremath{ m_t(m_t)^{\msbar}_{{\mathrm{SM}}}}}
\newc{\mtmtsmdrbar}{\ensuremath{ m_t(m_t)^{\drbar}_{{\mathrm{SM}}}}}
\newc{\mtmtmssmdrbar}{\ensuremath{ m_t(m_t)^{\drbar}_{{\mathrm{SUSY}}}}}
\newc{\mbmbmsbar}{\ensuremath{ m_b(m_b)^{\msbar} }}
\newc{\mbmbsmmsbar}{\ensuremath{ m_b(m_b)^{\msbar}_{{\mathrm{SM}}}}}
\newc{\mbmzsmmsbar}{\ensuremath{ m_b(\mz)^{\msbar}_{{\mathrm{SM}}}}}
\newc{\mbmzsmdrbar}{\ensuremath{ m_b(\mz)^{\drbar}_{{\mathrm{SM}}}}}
\newc{\mbmzmssmdrbar}{\ensuremath{ m_b(\mz)^{\drbar}_{{\mathrm{SUSY}}}}}
\newc{\mtaumzsmmsbar}{\ensuremath{ m_{\tau}(\mz)^{\msbar}_{{\mathrm{SM}}}}}
\newc{\mtaumzsmdrbar}{\ensuremath{ m_{\tau}(\mz)^{\drbar}_{{\mathrm{SM}}}}}
\newc{\mtaumzmssmdrbar}{\ensuremath{ m_{\tau}(\mz)^{\drbar}_{{\mathrm{SUSY}}}}}
\newc{\alphasmzms}{\ensuremath{\alpha_s(M_Z)^{\overline{MS}}}}
\newc{\alphaimzms}[1]{\ensuremath{\alpha_{#1}(M_Z)^{\overline{MS}}}}
\newc{\alphaemmz}{\ensuremath{\alpha_{\mathrm{em}}(M_Z)^{\overline{MS}}}}

\newc{\mzero}{\ensuremath{{m_0}}}
\newc{\mhalf}{\ensuremath{ m_{1/2}}}
\newc{\tanb}{\ensuremath{\tan\beta}}
\newc{\azero}{\ensuremath{ A_0}}
\newc{\bzero}{\ensuremath{ B_0}}
\newc{\signmu}{\ensuremath{\rm{sgn}\,\mu}}
\newc{\mueff}{\ensuremath{\mu_{\rm{eff}}}}
\newc{\lam}{\ensuremath{{\lambda}}}
\newc{\kap}{\ensuremath{{\kappa}}}
\newc{\alam}{\ensuremath{{A_{\lambda}}}}
\newc{\akap}{\ensuremath{{A_{\kappa}}}}
\newc{\hs}{\ensuremath{ H_s}}      
\newc{\mhs}{\ensuremath{ m_{H_s}}} 
\newc{\mgut}{\ensuremath{ M_{\rm GUT}}}
\newc{\mplanck}{\ensuremath{ M_{\rm P}}}      \newc{\mpl}{\ensuremath{ M_{\rm Pl}}}
\newc{\msusy}{\ensuremath{ M_{\rm SUSY}}}      \newc{\ms}{\ensuremath{ M_{\rm S}}}
 \newc{\mhl}{\ensuremath{m_\hl}} 
 \newc{\mhone}{\ensuremath{m_{h_1}}} 
 \newc{\mhtwo}{\ensuremath{m_{h_2}}} 
 \newc{\mglu}{\ensuremath{m_{\tilde g}}} 
 \newc{\mul}{\ensuremath{m_{\tilde{u}_L}}} 
 \newc{\mtone}{\ensuremath{m_{\tilde{t}_1}}} 
 \newc{\ma}{\ensuremath{m_A}} 
 \newc{\maone}{\ensuremath{m_{a_1}}} 
 \newc{\matwo}{\ensuremath{m_{a_2}}}
 \newc{\hone}{\ensuremath{h_1}}
 \newc{\htwo}{\ensuremath{h_2}}
 \newc{\aone}{\ensuremath{a_1}}
 \newc{\atwo}{\ensuremath{a_2}}
 \newc{\mhu}{\ensuremath{ m_{H_u}}}       
 \newc{\mhd}{\ensuremath{ m_{H_d}}}
 \newc{\mhusq}{\ensuremath{ m_{H_u}^2}}       
 \newc{\mhdsq}{\ensuremath{ m_{H_d}^2}}
 \newc{\mhuew}{\ensuremath{ m^{\ast}_{H_u}}}       
 \newc{\mhdew}{\ensuremath{ m^{\ast}_{H_d}}}
 \newc{\mhuewsq}{\ensuremath{ m^{\ast\, 2}_{H_u}}}       
 \newc{\mhdewsq}{\ensuremath{ m^{\ast\, 2}_{H_d}}}
 \newc{\hu}{\ensuremath{ H_u}}       
 \newc{\hd}{\ensuremath{ H_d}}
 \newc{\barmhu}{\ensuremath{ \bar{m}_{H_u}}}
 \newc{\barmhd}{\ensuremath{ \bar{m}_{H_d}}}

 \newc{\mqthree}{\ensuremath{m_{\widetilde{Q}_3}^2}}
 \newc{\muthree}{\ensuremath{m_{\tilde{u}_3}^2}}
 \newc{\mdthree}{\ensuremath{m_{\tilde{d}_3}^2}}
 \newc{\mlthree}{\ensuremath{m_{\widetilde{L}_3}^2}}
 \newc{\methree}{\ensuremath{m_{\tilde{e}_3}^2}}
 \newc{\mqtwo}{\ensuremath{m_{\widetilde{Q}_2}^2}}
 \newc{\mutwo}{\ensuremath{m_{\tilde{u}_2}^2}}
 \newc{\mdtwo}{\ensuremath{m_{\tilde{d}_2}^2}}
 \newc{\mltwo}{\ensuremath{m_{\widetilde{L}_2}^2}}
 \newc{\metwo}{\ensuremath{m_{\tilde{e}_2}^2}}
 \newc{\mqone}{\ensuremath{m_{\widetilde{Q}_1}^2}}
 \newc{\muone}{\ensuremath{m_{\tilde{u}_1}^2}}
 \newc{\mdone}{\ensuremath{m_{\tilde{d}_1}^2}}
 \newc{\mlone}{\ensuremath{m_{\widetilde{L}_1}^2}}
 \newc{\meone}{\ensuremath{m_{\tilde{e}_1}^2}}
 \newc{\msmul}{\ensuremath{m_{\tilde{\mu}_L}}}
 \newc{\msmur}{\ensuremath{m_{\tilde{\mu}_R}}}
 \newc{\msneumu}{\ensuremath{m_{\tilde{\nu}_{\mu}}}}
 \newc{\mone}{\ensuremath{M_1}}
 \newc{\monesq}{\ensuremath{M_1^2}}
 \newc{\mtwo}{\ensuremath{M_2}}
 \newc{\mtwosq}{\ensuremath{M_2^2}}
 \newc{\mthree}{\ensuremath{M_3}}
 \newc{\mthreesq}{\ensuremath{M_3^2}}
 \newc{\atau}{\ensuremath{{A_{\tau}}}}
 \newc{\at}{\ensuremath{{A_{t}}}}
 \newc{\ab}{\ensuremath{{A_{b}}}}
 \newc{\atausq}{\ensuremath{{A_{\tau}^2}}}
 \newc{\atsq}{\ensuremath{{A_{t}^2}}}
 \newc{\absq}{\ensuremath{{A_{b}^2}}}

 \newc{\dmzero}{\ensuremath{\Delta{_{m_0}}}}
 \newc{\dmhalf}{\ensuremath{\Delta{_{m_{1/2}}}}}
 \newc{\dmu}{\ensuremath{\Delta{_{\mu}}}}

 \newc{\pten}{\ensuremath{\psi_{10}}}
 \newc{\ffive}{\ensuremath{\phi_{5}}}
 \newc{\hfive}{\ensuremath{h_{5}}}
 \newc{\hbfive}{\ensuremath{h_{\bar{5}}}}
 \newc{\thet}{\ensuremath{\theta_{50}}}
 \newc{\thetb}{\ensuremath{\theta_{\,\overline{50}}}}
 \newc{\ptenhat}{\ensuremath{\hat{\psi}_{10}}}
 \newc{\ffivehat}{\ensuremath{\hat{\phi}_{5}}}
 \newc{\hfivehat}{\ensuremath{\hat{h}_{5}}}
 \newc{\hbfivehat}{\ensuremath{\hat{h}_{\bar{5}}}}
 \newc{\thethat}{\ensuremath{\hat{\theta}_{50}}}
 \newc{\thetbhat}{\ensuremath{\hat{\theta}_{\,\overline{50}}}}
 \newc{\si}{\ensuremath{\Sigma}}
 \newc{\mfive}{\ensuremath{m_5^2}}
 \newc{\mten}{\ensuremath{m_{10}^2}}
 \newc{\dfive}{\ensuremath{\Delta^2_5}}
 \newc{\dbfive}{\ensuremath{\Delta^2_{\bar{5}}}}
 \newc{\dfifty}{\ensuremath{\Delta^2_{50}}}
 \newc{\dfiftyb}{\ensuremath{\Delta^2_{\,\overline{50}}}}
 \newc{\msi}{\ensuremath{m_{\Sigma}^2}}
 \newc{\lamh}{\ensuremath{\lambda_{H}}}
 \newc{\lamhb}{\ensuremath{\lambda_{\bar{H}}}}
 \newc{\ah}{\ensuremath{A_{H}}}
 \newc{\ahb}{\ensuremath{A_{\bar{H}}}}
 \newc{\lams}{\ensuremath{\lambda_{S}}}
 \newc{\as}{\ensuremath{A_{S}}}
 \newc{\lamsig}{\ensuremath{\lambda_{\si}}}
 \newc{\asig}{\ensuremath{A_{\si}}}

 \newc{\msten}{\ensuremath{m_{16}^2}}
 \newc{\mhun}{\ensuremath{m_{126}^2}}
 \newc{\mhunb}{\ensuremath{m_{\bar{126}}^2}}
 \newc{\mthun}{\ensuremath{m_{210}^2}}
 \newc{\ahun}{\ensuremath{A_{\bar{126}}}}
 \newc{\yhun}{\ensuremath{Y_{\bar{126}}}}
 \newc{\aten}{\ensuremath{A_{10}}}
 \newc{\yten}{\ensuremath{Y_{10}}}
 \newc{\alone}{\ensuremath{A_{\lambda_1}}}
 \newc{\altwo}{\ensuremath{A_{\lambda_2}}}
 \newc{\althree}{\ensuremath{A_{\lambda_3}}}
 \newc{\althreeb}{\ensuremath{A_{\bar{\lambda_3}}}}
 \newc{\lone}{\ensuremath{\lambda_1}}
 \newc{\ltwo}{\ensuremath{\lambda_2}}
 \newc{\lthree}{\ensuremath{\lambda_3}}
 \newc{\lthreeb}{\ensuremath{\bar{\lambda_3}}}

\newc{\sigsip}{\ensuremath{\sigma^{\rm SI}_{p}}}	\newc{\sigsin}{\ensuremath{\sigma^{\rm SI}_{n}}}
\newc{\sigsdp}{\ensuremath{\sigma^{\rm SD}_{p}}}	\newc{\sigsdn}{\ensuremath{\sigma^{\rm SD}_{n}}}
\newc{\sigsi}{\ensuremath{\sigma^{\rm SI}}}	\newc{\sigsd}{\ensuremath{\sigma^{\rm SD}}}
\newc{\sigv}{\ensuremath{\sigma v}}
\newc{\abund}{\ensuremath{ \Omega h^2}}
\newc{\omegadm}{\ensuremath{ \Omega_{{\rm DM}}}}     \newc{\abunddm}{\ensuremath{ \Omega_{{\rm DM}} h^2}} 
\newc{\omegam}{\ensuremath{ \Omega_{{\rm m}}}}       \newc{\abundm}{\ensuremath{ \Omega_{{\rm m}} h^2}}
\newc{\omegab}{\ensuremath{ \Omega_{{\rm b}}}}	\newc{\abundb}{\ensuremath{ \Omega_{{\rm b}} h^2}}
\newc{\omegatot}{\ensuremath{ \Omega_{{\rm TOT}}}}
\newc{\omegacdm}{\ensuremath{ \Omega_{{\rm CDM}}}}   \newc{\abundcdm}{\ensuremath{ \Omega_{{\rm CDM}} h^2}}
\newc{\omegalambda}{\ensuremath{ \Omega_{\Lambda}}} \newc{\abundlambda}{\ensuremath{ \Omega_{\Lambda} h^2}}
\newc{\omegarad}{\ensuremath{ \Omega_{{\rm rad}}}}  \newc{\abundrad}{\ensuremath{ \Omega_{{\rm rad}} h^2}}
\newc{\rhocrit}{\ensuremath{ \rho_{\rm crit}}}
\newc{\rhochi}{\ensuremath{ \rho_{\chi}}}
\newc{\abunchi}{\ensuremath{\Omega_\chi h^2}}
\newc{\abundlsp}{\ensuremath{\Omega_{\rm LSP}h^2}}

\newc{\amu}{\ensuremath{ a_{\mu}}}        \newc{\amususy}{\ensuremath{ a_{\mu}^{\mathrm{SUSY}}}}
\newc{\amuexpt}{\ensuremath{ a_{\mu}^{\mathrm{expt}}}}        \newc{\amusm}{\ensuremath{ a_{\mu}^{\mathrm{SM}}}}
\newc\deltaamu{\ensuremath{\Delta a_{\mu}}} \newc{\deltaamususy}{\ensuremath{\delta a_{\mu}^{\mathrm{SUSY}}}}
\newc\gmtwo{\ensuremath{ (g-2)_{\mu}}} 
\newc{\deltagmtwomususy}{\ensuremath{\delta\left(g-2\right)_{\mu}^{\mathrm{SUSY}}}}
\newc{\deltagmtwomu}{\ensuremath{\delta\left(g-2\right)_{\mu}}}
\newc\BR{\ensuremath{\rm BR}}
\newc\bsgamma{\ensuremath{ b\rightarrow s \gamma }}
\newc\bxsgamma{\ensuremath{\overline{B}\rightarrow X_{s}\gamma}}
\newc\brbsgamma{\ensuremath{\BR\left(\bsgamma\right)}}
\newc\brbxsgamma{\ensuremath{\BR\left(\bxsgamma\right)}}
\newc\bsmumu{\ensuremath{B_s\to\mu^+\mu^-}}
\newc\brbsmumu{\ensuremath{\BR\left(B_s\to\mu^+\mu^-\right)}}
\newc\bdmmumu{\ensuremath{\overline{B}_d\to\mu^+\mu^-}}
\newc\bbbarmix{\ensuremath{\overline{B}_s\mbox{-}B_s}}      
\newc\delmbs{\ensuremath{\Delta M_{B_s}}}
\newc{\butaunu}{\ensuremath{B_u \rightarrow \tau \nu}}
\newc{\brbutaunu}{\ensuremath{\BR\left(B_u \rightarrow \tau \nu\right)}}

\newcommand*{\reffig}[1]{Figure~\ref{#1}}
 
        \newcommand*{\refeq}[1]{Equation~(\ref{#1})}

\newcommand*{\pythia}{\text{PYTHIA}}

\let\oldcite\cite
\renewcommand*{\cite}{~\oldcite}

\newcommand*{\madgr}{\texttt{MadGraph5\_aMC@NLO}}

\usepackage{calc,rotating}
\usepackage[english]{babel}

\usepackage{hyperref}
\usepackage[labelformat=simple]{subfig}

\usepackage{float}
\usepackage{amssymb}
\usepackage{amsthm}
\usepackage{latexsym}
\usepackage{dcolumn}

\Title{Blind Spots for Direct Detection with Simplified DM Models and the LHC}

\Author{Arghya Choudhury $^{1,2,*,\dagger,\ddagger}$, Kamila Kowalska $^{3,4,\ddagger}$, Leszek Roszkowski $^{1,4,\ddagger}$ , Enrico Maria Sessolo $^{3,4,\ddagger}$ and Andrew J. Williams $^{4,\ddagger}$}

\address{%
$^{1}$ \quad Consortium for Fundamental Physics, Department of Physics and Astronomy, University of Sheffield, Sheffield S3 7RH, United Kingdom\\
$^{2}$ \quad Consortium for Fundamental Physics, Department of Physics and Astronomy, University of Manchester, Manchester, M13 9PL, United Kingdom\\
$^{3}$ \quad  Fakult\"at f\"ur Physik, TU Dortmund, Otto-Hahn-Str.4, D-44221 Dortmund, Germany\\
$^{4}$ \quad  National Centre for Nuclear Research,  Ho{\. z}a 69, 00-681 Warsaw, Poland
\vspace{+2mm}}

\corres{Correspondence: a.choudhury@sheffield.ac.uk}

\firstnote{Speaker at ``Varying Constants and Fundamental Cosmology" -- VARCOSMOFUN'16} 

\secondnote{Email: a.choudhury@sheffield.ac.uk, kamila.kowalska@tu-dortmund.de, L.Roszkowski@sheffield.ac.uk, enrico.sessolo@ncbj.gov.pl, andrew.williams.2009@live.rhul.ac.uk}

\vspace{1mm}

\abstract{
Using the existing simplified model framework, we build several dark matter 
models which have suppressed spin-independent scattering cross section.~We show that the scattering cross section can vanish due to 
interference effects with models obtained by 
simple combinations of simplified~models.~For weakly interacting 
massive particle (WIMP) masses $\gtrsim$10 $\gev$, collider limits 
are usually much weaker than the direct detection limits coming 
from LUX or XENON100. 
However, for our model combinations, LHC analyses are 
more competitive for some parts of the parameter space.~The~regions with direct detection blind spots can be 
strongly constrained from the complementary use of several 
 Large Hadron Collider (LHC) searches like mono-jet, jets~+~missing transverse energy,  
heavy vector resonance searches, etc.~We evaluate the strongest limits for combinations of 
scalar~+~vector, ``squark'' + vector, and scalar + ``squark'' mediator, 
and present the  LHC 14 TeV projections. }


\begin{document}


\section{\label{sec:intro}Introduction}

Simplified model spectra (SMS)\cite{Goodman:2011jq,Abdallah:2014hon, 
Malik:2014ggr,Abdallah:2015ter,Abercrombie:2015wmb} have been one of 
the most popular frameworks for the interpretation of the bounds  
from mono-photon/mono-jet searches on direct production of dark 
matter (DM) at the Large Hadron Collider (LHC).~In such scenarios, cross section can be parametrized in terms of a 
few parameters, like the mediator mass 
or the couplings of the  dark matter with the visible sector. 
Early LHC results were often presented in the effective field theory 
(EFT) framework, which is a~good approximation as long as the mediator 
masses are well above the collision energy.~Various recent studies that analysed the direct detection (DD) and 
LHC bounds on dark matter have used simplified model scenarios (see, e.g.,~\cite{Abdallah:2014hon,Abdallah:2015ter}) 
for this reason and found that in general, for weakly interacting 
massive particle (WIMP) masses above $\sim$5 $\gev$ 
the limits on spin-independent DM-nucleon cross section (\sigsip) 
coming from mono-jet/mono-photon searches are not competitive with the 
limits from direct detection experiments like LUX\cite{Akerib:2013tjd} or XENON100\cite{Aprile:2012nq}.

On the other hand, several effects like cascade decays, or  
cancellations in the couplings, or~the~interference between 
different diagrams, which can produce ``blind spots'' for direct detection 
searches are not present in the most simple SMS. 
Thus, the detection issues would be interesting 
with the models that are halfway between a UV complete model and those SMS. 
In this work, we~combine three popular SMS in pairs.~We only consider the simplified model of DM represented by 
scalar mediators, colored scalar mediators 
and vector mediators  for which the  \sigsip\ is significant. In this study 
we only consider a Dirac fermion dark matter.

In this proceedings, we report on our recent paper\cite{Choudhury:2015lha} in 
which we  dedicated special attention to the direct detection blind spots, which arise  
from interference between different diagrams in ``less~simplified'' model frameworks (LSMS).

\section{The Model Blocks\label{sec:models}}

In this work we present a phenomenological analysis of a three 
``less simplified'' model (LSMS) of~DM. These LSMS, to some 
extent, mimic  the properties of more generic UV models.
\begin{itemize}
\item \textbf{Model 1.} 
Combining Higgs portal and vector mediators;

\item \textbf{Model 2.} Combining $t$-channel scalar mediators (charged under color) and Higgs portal;

\item \textbf{Model 3.} Combining  $t$-channel scalar mediators (charged under color) and vector mediators.
\end{itemize}	

\begin{figure}[H]
\centering
\includegraphics[width=0.9\textwidth]{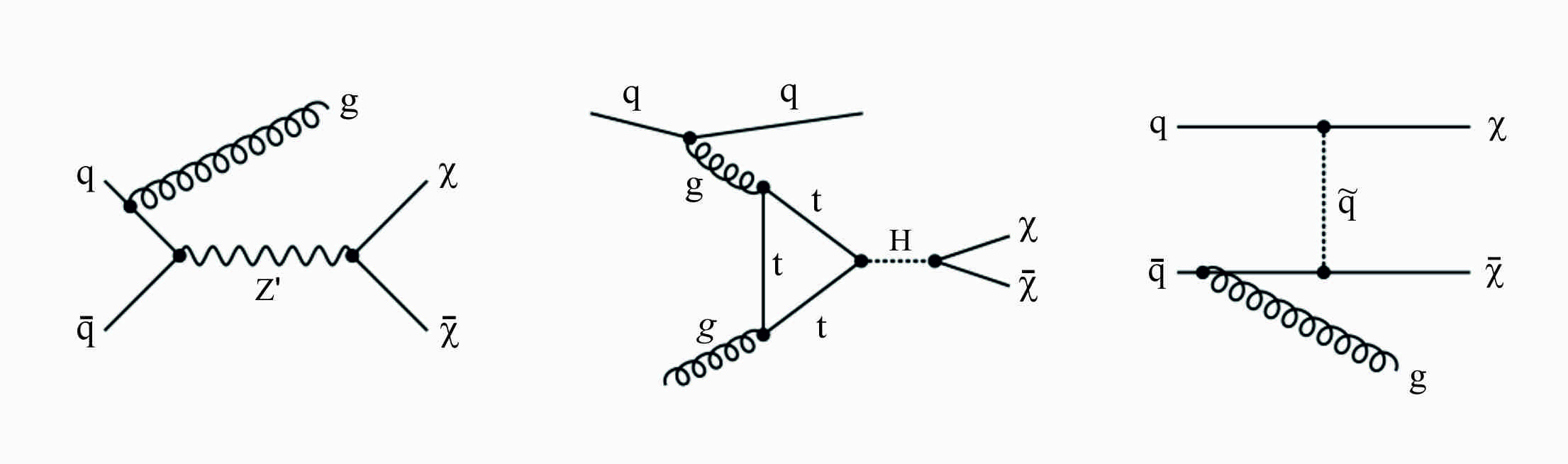}
\caption{\textls[-15]{Example of Feynman 
diagrams for $Z'$, Higgs and squark mediators which 
provide contributions} for monojet signature.}
\label{fig:monojet}
\end{figure}

For these three models, we always consider a Dirac fermion singlet DM. In the next subsections, we briefly recall the characteristics of the three above-mentioned SMS.\footnote{Example of Feynman diagrams for $Z'$, Higgs and squark mediators which 
provide contributions for monojet signature are given in Figure  \ref{fig:monojet}.}

\subsection{Vector Mediator\label{sec:vecmed}}

We consider a leptophobic $Z'$  mediator which has negligible 
mixing with the SM $Z$ boson. 
To~evade the stringent bounds from LHC di-lepton resonance 
searches, it has been assumed that $Z'$ also does not couple to 
the Standard Model leptons.

The relevant interaction terms for DM phenomenology at collider 
and direct detection experiments~are: 
\be\label{veclagr}
{\cal L} \supset Z'_\mu \bar{\chi}\gamma^\mu (g_\chi^V - g_\chi^A \gamma_5) \chi + \sum_i Z'_\mu \bar{q}_i\gamma^\mu (g_q^V - g_q^A \gamma_5) q_i\,,
\ee
where $g_q^V$, $g_q^A$, $g_\chi^V$, $g_\chi^A$ are universal vector quark coupling, 
universal axial-vector quark coupling, vector~and axial-vector couplings to the 
dark matter respectively.~In this study we restrict ourselves to 
vector boson exchange ($g_{\chi/q}^A$ = 0),~which contributes \sigsip.~Hence the free parameters become---
$\mchi,\,m_{Z'},\,g_{\chi}^V$ and $g_{q}^V$;  
where $\mchi$ is the DM mass and $m_{Z'}$ is the $Z'$ mediator mass.~As the product of $g^{V}_{\chi}$ and $g^{V}_q$ only matters 
for \sigsip, one may reduce the number of free parameters  to 3 
by assuming $g^{V}_{\chi} = g^{V}_q$.

\subsection{Higgs Portal/Scalar Mediator}
In this model, it is assumed that the fermion DM singlet ($\chi$) 
couples to a new singlet real scalar ($s$). 
The relevant terms for DM phenomenology are:\vspace{-3pt}
\be\label{hportal}
{\cal L} \supset - y_\chi \bar{\chi}\chi s - \mu_s s |\Phi|^2 - \lambda_s s^2 |\Phi|^2,
\ee
where $y_\chi$ is the Yukawa coupling between the singlet and the DM.  
Mixing between the scalar ($s$) and the SM Higgs doublet ($\Phi$) is induced by 
the mass term---
$\mu_s$.~After electroweak (EW) symmetry breaking, 
 one can diagonalize the mass matrix by a mixing matrix parametrized by a 
 mixing angle $\theta$. Then the relevant interaction terms in the Lagrangian become:\vspace{-3pt}
\be\label{higport}
{\cal L} \supset -y_\chi \left(h_{\textrm{SM}}\sin\theta+ H\cos\theta\right)\bar{\chi}\chi - 
\frac{1}{\sqrt{2}} \left(h_{\textrm{SM}}\cos\theta - H\sin\theta \right) \sum_f y_f \bar{f}f\,,\vspace{-6pt}
\ee 
where 
$f,\bar{f}$ are SM fermions and 
$y_f$ are the SM Yukawa couplings. 
The DM couples to quarks via the 
 heavy scalar mediator ($H$),  
  as well as the SM Higgs, $h_{SM}$.
  Thus, this type of models are  characterized by 
4 parameters---
$m_\chi,\,m_{H},\,\sin 2\theta,\,y_{\chi}$.

\subsection{Scalar $t$-Channel Mediators}
Finally, we consider  
scalar colored mediators which couple 
 the DM directly to the SM quarks.  
These scalar mediators are exchanged in the $t$-channel for DM production at the LHC.~Although our model is not based on SUSY, we adopt the  
notation $\tilde{q}$ which is used to represent  the 
squarks in the Minimal Supersymmetric Standard Model (MSSM). 
These  new scalars are charged under color and~flavor. 
We also assume that masses and couplings for first two generations 
are universal and the third generation squarks are beyond the reach of LHC.

The relevant terms for DM phenomenology are:\vspace{-3pt}
\be\label{squarks}
{\cal L} \supset \sum_{i=1,2} g_{\tilde{q}}\left( \tilde{u}^{\dag}_{i,R} \bar{\chi}P_R u_i + \tilde{u}^{\dag}_{i,L} \bar{\chi}P_L u_i + \tilde{d}^{\dag}_{i,R} \bar{\chi}P_R d_i + \tilde{d}^{\dag}_{i,L} \bar{\chi}P_L d_i   \right)+\textrm{ h.c.}, \vspace{-3pt}
\ee
where 
$\tilde{u}_{i,L(R)}$, $\tilde{d}_{i,L(R)}$, $u_i$ ($d_i$), $P_L$ ($P_R$) and 
$g_{\tilde{q}}$  are 
the~$i$th generation left (right) up-type squarks, 
the~$i$th generation left (right) down-type squark, 
the $i$th generation up (down) quarks, 
the left (right) chiral projection operators and 
the coupling strength respectively.~The stability of dark matter is assumed to be 
protected by  a discrete symmetry like R-parity.~This type of model is characterized by 3~parameters---
$ m_\chi,\,m_{\tilde{q}}$ and 
$ g_{\tilde{q}}$,~where $m_{\tilde{q}}$ is 
the universal squark mass and $g_{\tilde{q}}$ is the universal DM-squark~coupling.

\section{Methodology and Analysis of the Combined Models\label{sec:combos}}

Models 1--3 have been implemented using \texttt{FeynRules}
\cite{Alloul:2013bka}.~We have calculated \sigsip\ using $\tt micrOMEGAs\ v.4.1.8$\cite{Belanger:2013oya}. 
For event generation we have used 
\madgr\cite{Alwall:2014hca} and \texttt{\pythia8}\cite{Sjostrand:2007gs}. 
For Model 2--3, 
we have generated $\chi\bar{\chi}$ + jets, $\chi \tilde{q}$ + jets 
associated production, and $\tilde{q}\tilde{q}^{\ast}$ + jets. 
To obtain the exclusion limits from jets + missing energy 
and mono jet searches
 we have used \texttt{CheckMATE}\cite{Drees:2013wra} 
and our own codes which were previously used in 
References\cite{Kowalska:2015zja,Chakraborti:2014gea,Chakraborti:2015mra}.~For 14 TeV projections with jets~+~missing energy and mono-jet analysis 
we have followed the prescribed cuts given in References~\cite{ATL-PHYS-PUB-2014-007,ATL-PHYS-PUB-2014-010}. We~have also considered the limits on production cross section 
times branching ratio for  $Z'\rightarrow q\bar{q}$ and $Z'\rightarrow t\bar{t}$ 
from LHC 8 TeV data with 20.3 ${fb}^{-1}$ 
\cite{Aad:2015fna,Aad:2014aqa,ATLAS:2015nsi} and the corresponding 14\tev\ projections with $300~{fb}^{-1}$\cite{ATL-PHYS-PUB-2013-003}.~We have calculated the partial width of the 125 GeV Higgs to dark matter particles, 
$\Gamma_{h_{\textrm{SM}}\rightarrow \chi\bar{\chi}}$\,, 
using \texttt{CalcHEP}\cite{Belyaev:2012qa} to compare with the limit from\cite{ATLAS-CONF-2015-044}. 
Considering all these LHC searches, we have compared the upper limits on production cross section times branching ratio or the quantity ``upper limits on number of BSM events after 
all cuts ($N_{BSM}$)'' with our models for different values of coupling 
(e.g., $g_{\chi/q}^V,\,y_\chi$ etc.) to obtain the 
new limits.  
\subsection{Model 1: Combining \boldmath$Z'$ and Higgs Portal \label{sec:higz}} 
In the first LSMS (Model 1) we consider a $Z'$ vector boson 
and an extended scalar sector. Then~one~can write the relevant terms 
of Lagrangian as the sum of Equations (\ref{veclagr}) and (\ref{higport}).
With~the assumption that $g_{\chi}^V=g_{q}^V\equiv g_{\chi/q}^V$ 
the free parameters (6) become: 
$ \mchi,\,m_{Z'},\,m_H,\,\theta,\,y_\chi,\,g_{\chi/q}^V $. 

We assume mixing as maximal as is allowed by the LHC constraints, perturbativity of the couplings, and EW precision observables: $\theta=0.2$. The cross section \sigsip\ depends mildly on the angle, via $\sin 2\theta$.~More detailed analytical formulas for  
\sigsip\ can be found in Refernce\cite{Choudhury:2015lha}.

\begin{figure}[H]
\centering
\subfloat[]{%
\label{fig:a}%
\includegraphics[width=0.36\textwidth]{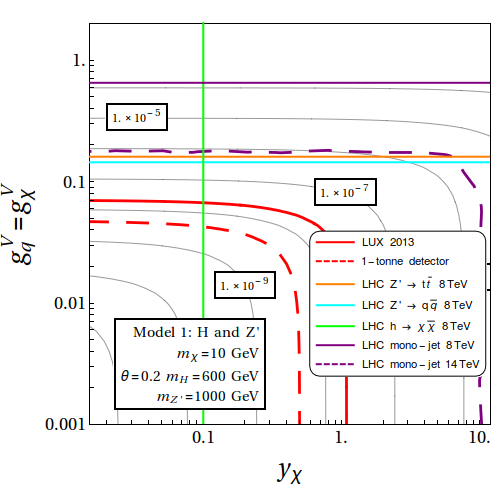}
}%
\subfloat[]{%
\label{fig:b}%
\includegraphics[width=0.36\textwidth]{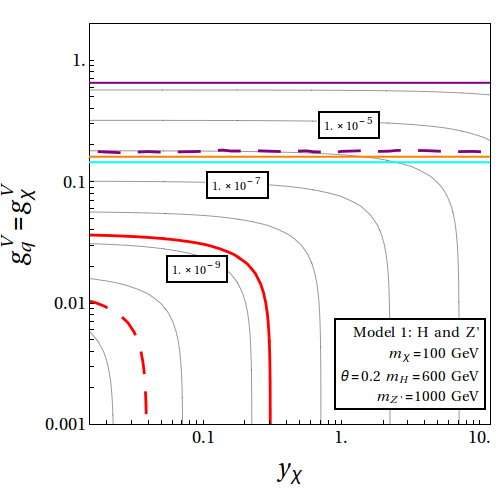}
}\\
\subfloat[]{%
\label{fig:c}%
\includegraphics[width=0.36\textwidth]{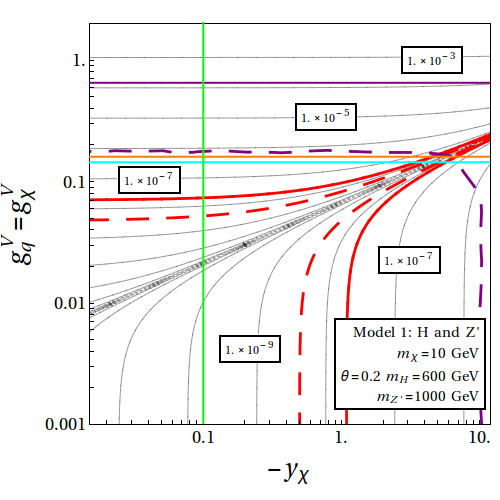}
}%
\subfloat[]{%
\label{fig:d}%
\includegraphics[width=0.36\textwidth]{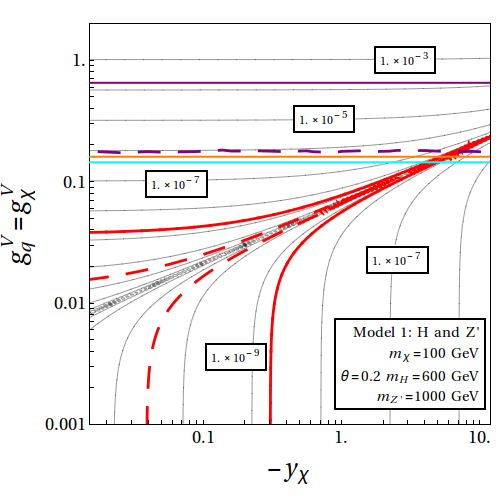}
}%
\caption{(\textbf{a}) \sigsip\ (pb) in the ($y_\chi$, $g_{\chi/q}^V$) plane for a Model~1.~We set  $\theta = 0.2$, $\mchi = 10\,\gev$, \mbox{$m_{Z'}=1000\,\gev$} and $m_H = 600\,\gev$. 
For other details see text; (\textbf{b}) Same as (\textbf{a}) with $\mchi=100\,\gev$; (\textbf{c}) Same as (\textbf{a}) but the sign of $y_{\chi}$ is negative; (\textbf{d}) Same as (\textbf{c}) but $\mchi=100\,\gev$.}
\label{fig:Z_H}
\end{figure}

In \reffig{fig:Z_H}a we present the contours of \sigsip\ in pb in the 
($y_{\chi}$, $g_{\chi/q}^V$) plane for $\mchi=10\,\gev$, 
\mbox{$m_H = 600\,\gev$} and $m_{Z'}=1000\,\gev$.
If $y_{\chi}>0$\,, $m_{H}\gg m_{h_{\textrm{SM}}}$, and $g_{\chi}^V=g_{q}^V$, destructive interference does not take place (see Equation (3.3)\footnote{The~cancellation can only happen in the nonrelativistic limit.
The~formula for differential WIMP-nucleus scattering cross section is given 
in Equation  (3.2) of Reference~\cite{Choudhury:2015lha} and for
the relativistic WIMP-quark scattering see Equation 3.4 of Reference~\cite{Choudhury:2015lha}.} 
 of Reference~\cite{Choudhury:2015lha}).~LUX~bounds and XENON-1T projected reach are shown by 
solid red and dashed red lines.~Solid (dashed) purple line presents the upper bound from 
8 TeV mono-jet searches (14~TeV~projected reach). 
The green solid vertical line provides the upper limit 
on the $y_{\chi}$ obtained from a ATLAS/CMS combined analysis 
of $\Gamma_{h_{\textrm{SM}}\rightarrow \chi\bar{\chi}}$\cite{ATLAS-CONF-2015-044}.
The limits from $Z'\rightarrow q\bar{q}$ and $Z'\rightarrow t\bar{t}$  
are presented by solid orange line and  solid cyan line.

\reffig{fig:Z_H} shows that the direct detection limits on coupling 
($g_{\chi/q}^V$) from LUX data is more severe than the collider limits 
in general. Only for $\mchi\lesssim 62\gev$, 
the invisible width of the 125 GeV Higgs boson 
(green line in \reffig{fig:Z_H}a,c)
is significantly more 
 constraining than the DD bounds.  
In \reffig{fig:Z_H}c,d, 
due to the choice 
$y_{\chi}<0$\, (or if it is positive but $g_{\chi}^V=-g_{q}^V$), 
one gets suppressed \sigsip. This happens due to the destructive 
interference of the diagrams corresponding to the Higgs portal 
$Z'$  (see~Equation~(3.3) of Reference\cite{Choudhury:2015lha}). 
The blind spot in the plots \reffig{fig:Z_H}c,d 
for for $\mchi=10\gev$ and 100 GeV respectively  is 
a narrow diagonal region, over which the value of \sigsip\ visibly drops below
the potential reach of tonne-scale detectors. 
The condition for the blind spot can be written as: 
\be 
y_{\chi}\approx -\left(\frac{8.22\times 10^7\textrm{ GeV}^2}{m_{Z'}^2}\right)
\frac{g_{\chi}^V g_{q}^V}{\sin 2\theta\, \left(1-\frac{m_{h_{\textrm{SM}}}^2}{m_H^2}\right)}\,.\label{BS1}
\ee

\begin{figure}[H]
\centering
\subfloat[]{%
\label{fig:a}%
\includegraphics[width=0.36\textwidth]{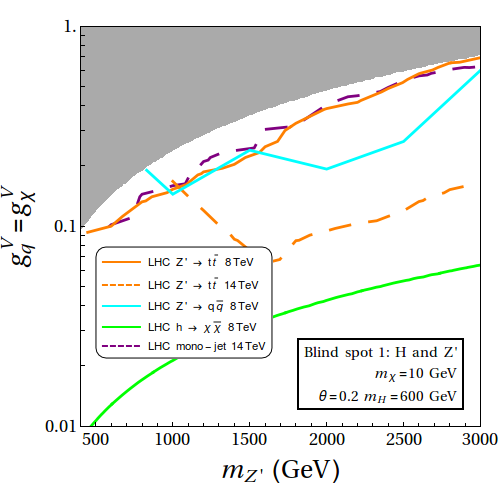}
}
\hspace{0.07\textwidth}
\subfloat[]{%
\label{fig:c}%
\includegraphics[width=0.36\textwidth]{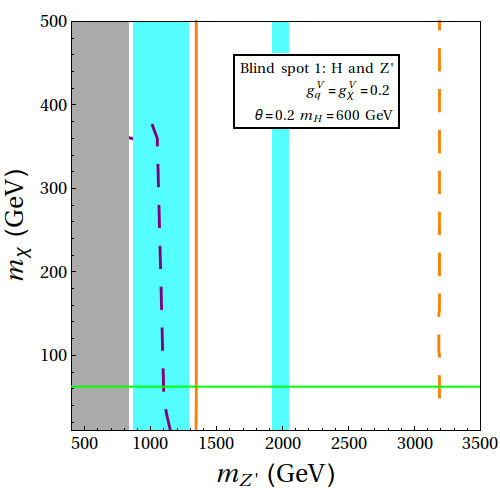}
}%
\caption{(\textbf{a}) Interplay of collider constraints for the blind spot regions parametrized by \refeq{BS1} in the ($m_{Z'}$, $g_{\chi/q}^V$) plane. Here $\mchi=10\,\gev$, $m_{H}=600\,\gev$, and $\theta=0.2$; 
(\textbf{b}) The bounds projected to the ($m_{Z'}$, \mchi) plane for $g_{\chi/q}^V=0.2$ and $m_{H}=600\,\gev$. 
See text for other details.}
\label{fig:Z_H_det}
\end{figure}

Equation~(\ref{BS1}) shows that the contributions to the amplitude of the diagrams 
from the Higgs portal and $Z'$ are of comparable size 
for comparable coupling strengths if $m_{Z'}$ is at least of 
the order of a \tev\ or larger. When Condition~(\ref{BS1}) 
is satisfied, Model~1 is beyond the reach of direct detection 
searches but it can be studied by collider means. 
For this blind spot region, we show the effect of 
 monojet searches, 125 GeV Higgs partial width measurements 
 and $Z'$ resonances searches in the ($m_{Z'}$, $g_{\chi/q}^V$) plane 
 in \reffig{fig:Z_H_det} for fixed $m_{H}=600\gev$ and two values 
 of DM mass: (a) $\mchi=10\gev$,  
(b) $\mchi=100\gev$. 

The grey regions at the top of 
\reffig{fig:Z_H_det}a are not allowed as 
$y_{\chi}$ becomes nonperturbative ($y_{\chi}> 4\pi$). 
Color coding in \reffig{fig:Z_H_det}a is same as in  
\reffig{fig:Z_H}.~The bounds from the direct $Z'$ resonance  searches 
and the single-jet searches remain almost unchanged 
over a large range of \mchi.  
In \reffig{fig:Z_H_det}, we observe that the 
limit from $Z'\rightarrow t\bar{t}$ searches at  8 TeV
(solid orange line), 
 $Z'\rightarrow q\bar{q}$ searches\cite{Aad:2014aqa} at  8 TeV  
(cyan~solid~line) and future projection of ATLAS mono-jet 
searches (dashed purple line) are comparable to each~other. 
For resonant searches, 14\tev\ data (dashed orange line) improved 
significantly on the 8\tev\ data for $Z'$ masses above 1500\gev. 
In \reffig{fig:Z_H_det}b (for fixed $g_{\chi/q}^V=0.2$), we show that 
mono-jet search is always weaker than the direct search for a $Z'$. 
\unskip

\subsection{Model 2: Combining Higgs Portal and Squarks}

Model 2 features some of the properties of SUSY models of DM. In particular 
the limits are similar to those cases where the neutralino couples to 
the SM Higgs and additional heavy Higgs bosons.  In~contrast with MSSM, the DM in our model is a Dirac fermion with free couplings, 
and the additional scalar is a SM singlet. 
The relevant terms of Lagrangian for Model~2 
is simply the sum of \mbox{Equations (\ref{higport}) and (\ref{squarks}}). 
Hence we have 6 free parameters for  Model~2---
$\mchi,\,m_{\tilde{q}},\,m_H,\,\theta,\,y_\chi, \, g_{\tilde{q}} $.

\begin{figure}[H]
\centering
\subfloat[]{%
\label{fig:a}%
\includegraphics[width=0.4\textwidth]{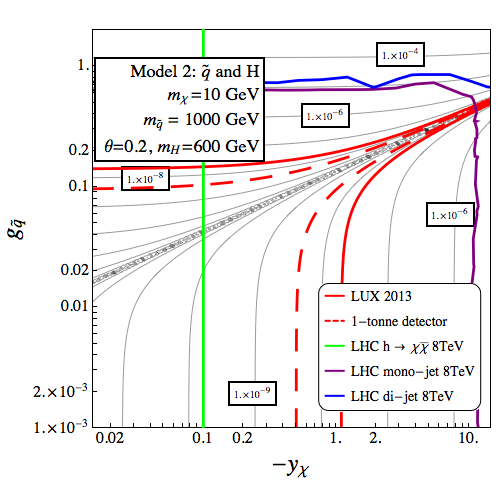}
}%
\hspace{0.07\textwidth}
\subfloat[]{%
\label{fig:b}%
\includegraphics[width=0.4\textwidth]{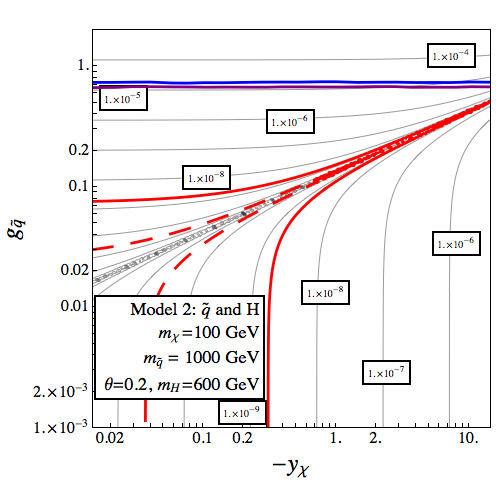}
}%
\caption{(\textbf{a})  \sigsip\ (pb) in the ($y_\chi$, $g_{\tilde{q}}$) plane 
for a Model~2. We have fixed $\mchi = 10\,\gev$, $m_{\tilde{q}} = 1000\,\gev$, 
$\theta = 0.2$, and $m_H=600\,\gev$. The full parameter space 
shown in this figure is within reach of 14\,\tev\ monojet and jets + missing energy 
searches; (\textbf{b}) Same as (\textbf{a}) but $\mchi = 100\,\gev$.}
\label{fig:squark_h_couplings}
\end{figure}

We present our results for Model 2 with fixed values 
of  $m_{\tilde{q}} = 1000\,\gev$,  $\theta = 0.2$, $m_H=600\,\gev$ 
and $\mchi = 10\,\gev$ (100 GeV) in 
\reffig{fig:squark_h_couplings}.  Similar to Model~1, 
cancellations in the amplitude for \sigsip\  
do not occur for $y_{\chi}>0$. 
Hence, we restrict ourselves for the 
case $y_\chi<0$ and one can write the 
blind spot condition for Model~2 as: 
\be 
y_{\chi}\approx -\left(\frac{2.05\times 10^7\textrm{ GeV}^2}{m_{\tilde{q}}^2-\mchi^2}\right)
\frac{g_{\tilde{q}}^2}{\sin 2\theta\, \left(1-\frac{m_{h_{\textrm{SM}}}^2}{m_H^2}\right)}\,.\label{BS2}
\ee

\reffig{fig:squark_h_couplings} a,b 
present the contours of \sigsip\ in the 
($y_\chi$, $g_{\tilde{q}}$) plane 
for $\mchi = 10\,\gev$ and $\mchi = 100\,\gev$ respectively. 
The color convention in \reffig{fig:squark_h_couplings} 
is exactly same as in \reffig{fig:Z_H}. 
Additionally,  the upper limit from the ATLAS 8\,\tev\ squarks search 
in jets + missing $E_T$\cite{Aad:2014wea} is presented by the solid blue line 
(see also\cite{Khachatryan:2015vra} for the CMS bound).
Similar to Model~1, 
the bound on $|y_{\chi}|$ from the 
invisible width of the 125 GeV Higgs is 
much stringent than the DD bounds for 
 $\mchi < 62.5\gev$ (see green solid line in \reffig{fig:squark_h_couplings}a).
This bound from the invisible width does not 
applicable to \reffig{fig:squark_h_couplings}b 
and the blind spot regions which are 
not in reach of underground DD experiments 
remain essentially unconstrained for $\mchi\gtrsim 62\,\gev$.~The dependence of the bounds on 
$m_{\tilde{q}}$ and $\mchi$ when  
\refeq{BS2} holds are presented in great details in\cite{Choudhury:2015lha}.

\subsection{Model 3: Combining $Z'$ and Squarks\label{sec:sqz}} 

We have designed  
Model~3  to mimic a UV completion 
characterized by an additional $U(1)_X$ symmetry
that remains unbroken down to collider energies (see, e.g.,\cite{Athanasopoulos:2014bba}).~Among the several possibilities, one way of building a gauge invariant LSMS 
with the squarks and $Z'$ mediated simplified models 
is the following, which allows the squarks to have the same coupling to the $Z'$ as the quarks, and could be seen as an approximation of a full UV theory involving an extended gauge symmetry and a supersymmetric sector (for details see Section 3.3 of Reference\cite{Choudhury:2015lha}).
Despite being apparently rather involved, the phenomenology of Model~3 
is represented by 6 free parameters---$\mchi, m_{\tilde{q}},\,m_{Z'},\, g_{\chi}^V,\, g_q^V,\, g_{\tilde{q}}$. 
With the additional assumption $g_{\chi}^V=\pm g_q^V\equiv g_{\chi/q}^V$, 
this number is further reduced to 5.

\begin{figure}[H]
\centering
\subfloat[]{%
\label{fig:a}%
\includegraphics[width=0.4\textwidth]{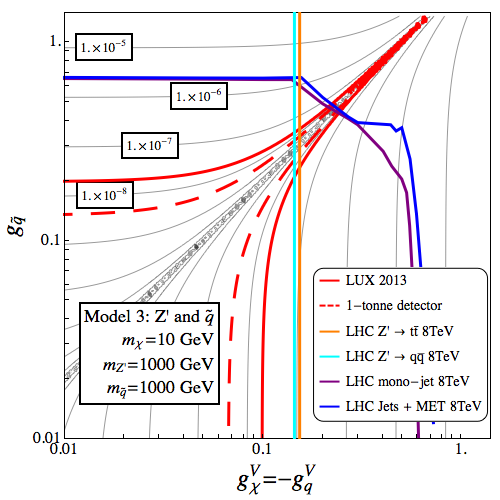}
}%
\hspace{0.07\textwidth}
\subfloat[]{%
\label{fig:b}%
\includegraphics[width=0.4\textwidth]{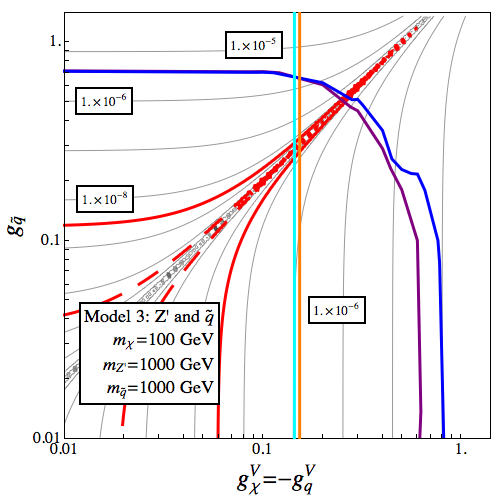}
}%
\caption{(\textbf{a}) 
\sigsip\ (pb)  in the ($g_{\chi}^V=-g_{q}^V$, $g_{\tilde{q}}$) plane 
for a combined Model 3 ($Z'+\textrm{squark}$ mediator simplified model). 
The masses are fixed at $\mchi = 10\,\gev$, $m_{Z'}=1000\,\gev$, and $m_{\tilde{q}}=1000\,\gev$; 
(\textbf{b})~Same as (\textbf{a}) but $\mchi=100\,\gev$.}
\label{fig:Z_squark}
\end{figure}

We show the \sigsip\ contours in  \reffig{fig:Z_squark}a 
for Model~3 in the  
($g_{\chi/q}^V$, $g_{\tilde{q}}$) plane for 
fixed $m_{\tilde{q}}=1000\,\gev$ and $m_{Z'}=1000\,\gev$. 
Color conventions are same as \reffig{fig:Z_H} or  
\reffig{fig:squark_h_couplings}.~The case with \mbox{$\mchi = 100\,\gev$} is shown in \reffig{fig:Z_squark}b.~It may be noted that the LHC limits barely move by 
changing the DM mass, but~direct detection limits reach their 
close-to-maximal strength when $\mchi = 100\,\gev$. The~\mbox{jets + missing} energy and mono-jet searches 
put comparable constraints on the couplings (blue and purple line). 
 The interplay of LHC limits on the mediators' mass for blind 
 spot regions are discussed in details in Section 3.3 
 of  Reference\cite{Choudhury:2015lha}. A complementary use of 
 different detection strategies is needed to constrain a large part 
 of the parameter space which is invisible in direct detection  experiments.

\section{\label{sec:summary}Summary and Conclusions}

In this work we have presented three dark matter models 
(LSMS) which are simple extensions of simplified DM models 
and mimic some properties of more realistic models 
without introducing an~excessively large number of parameters. 
We mainly focussed to the scenarios where the 
interference between different diagrams produces blind 
spot for direct detection experiments.~These blind spot regions are then further tested  
with current LHC limits and also with future 
LHC projections.

In general for $\mchi\geq10\gev$, the DD bounds on 
\sigsip\  excludes the coupling constants of 
WIMP by at least one order of magnitude more 
strongly than any of the LHC searches considered here. 
The~exceptions are: 
(i) In models with  a Higgs portal (Model 1 and Model 2), 
for $\mchi\lesssim 1/2\,m_{h_{\textrm{SM}}}$ the parameter 
space is also strongly constrained by the 
invisible width of the 125 GeV Higgs boson;  
(ii)~In~the parameter space corresponding to suppressed \sigsip\ due to 
interference effect, i.e., blind spot regions, 
 LHC searches can effectively place strong bounds,
especially at the end of Run~2.

We have found the following characteristics for the blind spot regions 
of Models 1--3:
\begin{itemize}
\item The Model 1 (combination of Higgs portal and $Z'$) 
is at present not constrained at all by mono-jet searches 
for the assumption $g_{\chi}^V=g_q^V$. 
Moreover, under this assumption, the future searches of heavy $Z'$ resonances 
at the LHC will be most effective to probe the blind spot regions. 

\item In Model 2 and Model 3 (involving squark-like mediators), 
the current limits  on the coupling 
$g_{\tilde{q}}$ from jets + missing energy and mono-jet searches 
are comparable. However, according to LHC 
future projections\cite{ATL-PHYS-PUB-2014-007,ATL-PHYS-PUB-2014-010}, 
the jets + missing energy searches at the 14\tev\ LHC  will 
outperform the expectations for mono-jet searches 
in the parameter space with blind spots. 
\end{itemize}

In general we find that, to constrain the DM models at the colliders, it is crucial 
to use the complementarity of different search strategies.
A lot of well motivated DM models, which are not necessarily 
constrained by DD bounds, demand careful attention at LHC or future colliders. 
In this work, we have investigated such scenarios in terms 
of a less-simplified model framework which can be explored at the LHC.

\vspace{6pt}





\acknowledgments{A.C. and L.R. are supported by the Lancaster-Manchester-Sheffield Consortium
for Fundamental Physics under STFC Grant No. ST/L000520/1. 
K.K. is supported in part by the DFG Research Unit FOR 1873 ``Quark Flavour Physics and
Effective Field Theories''. L.R. and E.M.S. are supported in part by the 
National Science Council (NCN) research grant No. 2015-18-A-ST2-00748. The work of E.M.S. is supported in part by the Alexander von Humboldt Foundation. 
The use of the CIS computer cluster at the National Centre for Nuclear Research in Warsaw is
gratefully acknowledged.}



\bibliographystyle{mdpi}


\bibliography{BF_12}

\end{document}